%Paper: hep-th/9205023
%From: Rue-Ron Hsu <NCKUT057%TWNMOE10.BITNET@pucc.princeton.edu>
%Date: Tue, 12 May 92 14:40:13 EST

\input phyzzx
\titlepage
\title{ Stability Analysis of a Stringy Black Hole}
\author{Rue-Ron Hsu, Green Huang and Wei-Fu Lin}
\address{Department of Physics, National Cheng Kung University,}
\address{Tainan, Taiwan, 70101, Republic of China}
\author{Chin-Rong Lee}
\address{Institute of Physics, National Chung-Cheng University,}
\address{Chia-Yi, Taiwan, 62117, Republic of China}
\vskip 0.0cm
\hskip 4.9 in { 97.60Lf}
\abstract
We investigate the stability of charged black holes in two-dimensional
heterotic
string theories that were recently discussed by McGuidan, Nappi
and Yost. In the framework of small time-dependent perturbation, we find that
these black holes are linearly stable.
\endpage

Recently, exact solutions of string theory in the form of two-dimensional black
holes have attracted much interest\refmark{1-4}. ~These stringy black holes
aris
from the non-minimal coupling of a scalar field, the dilaton, to gravity.
The presence of the dilaton may have important new effects which
should distinguish them from the known solutions in general relativity
\refmark{5}. ~The purpose of this letter is to analyze the behaviour of a
two-dimensional black hole under perturbations\refmark{5-8}. ~Our
results are, however, different from those of Ref.6 and Ref.8. The black hole
under study is linearly stable.

The starting point of our analysis is the string-inspired effective action---
McGuidan, Nappi and Yost (MNY) action\refmark{9} ~\foot {We shall use
the conventions given in Ref.[10]}
$$\eqalign{ S = & \int d^2x\sqrt{-G} e^{-2\phi}\bigl[R - 4(\nabla \phi)^2 - c -
	   {1\over 4}F^2\bigr],\cr}\eqno(1)$$
where $ F_{\mu\nu}$ is the Maxwell tensor and $\phi$ is the dilatin
field. The corresponding equations of motion are
$$\eqalign{R_{\mu\nu}= 2\phi_{;\mu;\nu} + {1\over2}F_{\mu\sigma}F^\sigma_{ \nu}
	,\cr ( e^{-2\phi}F^{\mu\nu} )_{;\nu} =	0,\cr
	R + 4\phi_{;\rho}\phi^{;\rho} - 4\phi^{;\rho}_{ ;\rho}	-c-{1\over 4}
	  F^2 =  0.\cr}\eqno(2)$$
Choosing the form( the Schwarzschild-like gauge\refmark{2} ) of the metric
$$\eqalign{ g_{\mu\nu} =& \pmatrix{ g(r) & 0  \cr
				   0	&   -{1\over {g(r)}}},\cr}$$
the solution is
$$\eqalign{ \phi = & \phi_0 - {Q\over 2}r, \cr
	    f = & \sqrt{2} Qqe^{-Qr}  ( = F_{tr}),\cr
	    g = & 1 - 2me^{-Qr} + q^2e^{-2Qr} ,\cr}\eqno(3)$$
where $\phi_0$ and $Q$ are integration constants and asymptotic flatness
requires central charge $c = -Q^2$ and $Q > 0$. The free parameters $m$ and $q$
proportional to the mass and the charge of the black hole respectively.
There are event horizons, in the case of $m^2 > q^2$,  at $r_{\pm}={1\over Q}
\ln {(m\pm \sqrt {m^2-q^2})}$. The asymptotical region
is at $r \to \infty $, while the singularity is at $r=- \infty $.

We now proceed to consider perturbations to the metric, dilaton, and Maxwell
tensor, following along the work of Chandrasekhar\refmark{11}. In the present
case, the perturbed metric can be chosen as
$$\eqalign{ ds^2 = & (g + \epsilon_1 )(dt + \epsilon_{\omega}dr )^2 -
	    ({1\over g} + \epsilon_2 ) dr^2,\cr}$$
where $\epsilon_1, \epsilon_2, \epsilon_{\omega}$ are functions of $r$ and $t$.
They are regarded as small variations. The perturbation leading to
non-vanishing value of $\epsilon_{\omega}$ induces a dragging of the inertial
frame while those leading to non-vanishing $\epsilon_1$ and $\epsilon_2$ have
no
such effect. The former is called axial perturbation and the latter are called
polar perturbations. As one may expect from these two different behaviour of
the fluctuations. They can be considered independent of each other. Since there
are no axial modes in the analysis of the first-order variations of the
two-dimensional black hole \refmark{6}, ~in the following analysis, we will
only
consider the polar perturbations. As for the variations of the dilaton
field and Maxwell tensor, we introduce the notation
$$\eqalign{ \phi (r) \to & \phi (r) + \epsilon_{\phi}(r,t),\cr
	    f(r) \to & f(r) + \epsilon_f (r,t),\cr}$$
where $\epsilon_{\phi}$ and $\epsilon_f$ are considered to be small
perturbations. Upon substituting these functions into the field equations
(eq.(2)) and keeping only up to linear terms in the perturbations, one
obtain the following five coupled differential equations:
$$\eqalign{& -{1\over 2}\sigma^2 g\epsilon_2 - {1\over 2}g\epsilon_{1,r,r} +
	  {1\over 4}g(g_{,r})^2\epsilon_2 + {1\over 4}g^2g_{,r}\epsilon_{2,r}
	  + {1\over 4}g_{,r}\epsilon_{1,r} - {1\over {4g}}(g_{,r})^2\epsilon_1
	  + 2\sigma^2\epsilon_\phi \cr
	   &\quad  + g\phi_{,r}\epsilon_{1,r}
	    + gg_{,r}\epsilon_{\phi,r} +gf\epsilon_f =	0,\cr}\eqno(4)$$
$$\eqalign{& -2i\sigma\epsilon_{\phi,r} + {1\over g}i\sigma
g_{,r}\epsilon_{\phi
	    + gi\sigma\phi_{,r}\epsilon_2 =  0,\cr}\eqno(5)$$
$$\eqalign{& \sigma^2{1\over {2g}}\epsilon_2 - {1\over
{4g^2}}g_{,r}\epsilon_{1,
	 + {1\over {2g}}\epsilon_{1,r,r} + {1\over {4g^3}}(g_{,r})^2\epsilon_1
	 - {1\over {2g^2}}g_{,r,r}\epsilon_1 - {1\over 4}g_{,r}\epsilon_{2,r} \cr
	 &\quad - {1\over {4g}}(g_{,r})^2\epsilon_2 - 2\epsilon_{\phi,r,r}
	   + g\phi_{,r}\epsilon_{2,r} + g_{,r}\phi_{,r}\epsilon_2 -
	   {1\over g}g_{,r}\epsilon_{\phi,r} - {1\over g}f\epsilon_f
	   + {1\over {2g^2}}f^2\epsilon_1 = 0,\cr}\eqno(6)$$
$$\eqalign{& \epsilon_f - {1\over {2g}}f\epsilon_1 - {1\over 2}gf\epsilon_2
	   - 2f\epsilon_{\phi} =  0,\cr}\eqno(7)$$
$$\eqalign{& -\sigma^2\epsilon_2 - \epsilon_{1,r,r}+{1\over
2}(g_{,r})^2\epsilon
	   + {1\over 2}gg_{,}\epsilon_{2,r} +gg_{,r,r}\epsilon_2 \cr
	 &\quad + {1\over {2g}}g_{,r}\epsilon_{1,r}-{1\over
{2g^2}}(g_{,r})^2\epsilon_1
	   + {1\over g}g_{,r,r}\epsilon_1 -8g\phi_{,r}\epsilon_{\phi,r}
	   + 4g^2(\phi_{,r})^2\epsilon_2 + {4\over g}\sigma^2\epsilon_{\phi} \cr
	 &\quad + 4g_{,r}\epsilon_{\phi,r} +4g\epsilon_{\phi,r,r}
	   - 6gg_{,r}\phi_{,r}\epsilon_2 - 2g^2\phi_{,r}\epsilon_{2,r}
	   + 2\phi_{,r}\epsilon_{1,r} - {2\over g}g_{,r}\phi_{,r}\epsilon_1 \cr
	 &\quad + f\epsilon_f - {1\over {2g}}f^2\epsilon_1 - {1\over 2}gf^2\epsilon_2
	   =  0, \cr}\eqno(8)$$
where we restrict ourselves to look only for solutions with the time dependence
$e^{i\sigma t}$.

Now, we readily obtain:
$$\eqalign{ g\phi_{,r}\epsilon_2 = & 2\epsilon_{\phi,r} -
	    {1\over g}g_{,r}\epsilon_{\phi} ,\cr}\eqno(9)$$
$$\eqalign{ \epsilon_f = {1\over {2g}}f\epsilon_1  +{1\over 2}gf\epsilon_2
	   + 2f\epsilon_{\phi} ,\cr}\eqno(10)$$
from eq.(5) and eq.(7) respectively. Eq.(6) multiplying by $2g$ and combining
with eq.(8) yields
$$\eqalign{ 2\sigma^2{1\over {g^2}}\epsilon_{\phi} - 2\epsilon_{\phi,r,r}
	    + \phi_{,r}({\epsilon_1\over g} + g\epsilon_2)_{,r} = & 0.\cr}
	   \eqno(11)$$
It is quite amazing, that eq.(4) and eq.(8) are identically
zero by using the equations of motion (eq.(2)), and eqs.(9)-(11).
This fact simplfies the analysis very much.
It turns out that there are four variables, $\epsilon_1, \epsilon_2,
{}~~\epsilon_{\phi}$ and $\epsilon_f$, but only three independent equtions,
 eqs.(9)-(11). Once
again, we have the freedom to pick up a constraint---gauge freedom. We work
under the Schwarzschild-like gauge (gauge fixing) so that
$$\eqalign{ {1\over g}\epsilon_1 + g\epsilon_2 = & 0. \cr}\eqno(12)$$
Substituting eq.(12) into eq.(11), we obtain, finally, for the amplitude
$\epsilon_{\phi}(r) $ in $\epsilon_{\phi}(r,t) = \epsilon_{\phi}(r)e^{i\sigma
t}
the following basic eigenvalue equation:
$$\eqalign{ g^2\epsilon_{\phi,r,r} = & \sigma^2\epsilon_{\phi}.\cr}\eqno(13)$$

The remaining procedure is now clear. It is very helpful to bring eq.(13)
into the form of a Schr$\ddot o$dinger-like equation. For this, we introduce
a new coordinate $r^*$ given by
$$\eqalign{ {{dr^*}\over {dr}} = & {1\over {g(r)}}.\cr}$$
In order to describe the physically interesting region which is between the
oute
event horizon $r_+$ and the asymptotic flat region as $r \to \infty$,
this new coordinate $r^*$ will take values from $-\infty$ to $+\infty$.
Also, the eigenvalue equation (13), then, becomes the form of a
Schr$\ddot o$diger equation:
$$\eqalign{\bigl[-{{d^2}\over
{dr^{*^2}}}+V(r^*)\bigr]U(r^*)=&(-\sigma^2)U(r^*),
	   \cr}\eqno(14)$$
where $\epsilon_{\phi} = g^{1\over 2} (r^*)U(r^*)$ and
$V(r^*) = {3\over {4g^2}}(g_{,r^*})^2 - {1\over {2g}} g_{,r^*,r^*}$.
The energy spectrum of eq.(14) determines the spectrum for $\sigma^2$. In
particular, positive energy bound states of the Schr$\ddot o$dinger equation
, in our case, correspond to exponentially growing modes. The potential
$V(r^*)$ are plotted in Fig.1. Since
the potential provide no bound state\refmark{12},
we conclude that the charged black holes are linearly stable.

Finally, we should point out that the black hole solution obtained by Witten
and by Mandal $\it et. al.$ \refmark{1-2} is a special case of the charged
black
hole, discussed by McGuidan, Nappi and Yost\refmark{9}, ~by switching off the
charges ($q=0$). So the neutral black hole is stable. Our results
are also consistent with the conclusions obtained recently
by de Alwis\refmark{13}. He showed that
the two dimensional stringy black hole solution is
semi-classically stable to tachyon fluctuation.
\vskip 2cm
\centerline{\bf Acknowledgements}
\vskip 1cm
This work is supported in part by the National Science Council of Republic of
China under Grant NSC 81-0208-M194-07 and NSC 81-0208-M006-05.
\endpage
\def\jo{\journal}
\def\prl{Phys. Rev. Lett.}
\def\pr{Phys. Rev.}

\def\mpl{Mod. Phys. Lett.}
\def\np{Nucl. Phys.}
\ref{E.~Witten\jo\pr&D44(91)314.}
\ref{G.~Mandal, A.~Sengupta and S.~Wadia\jo\mpl&A6(91)1685.}
\ref{S.~Elitzur, A.~Forge and E.~Rabinovici\jo\np&B359(1991)581.}
\ref{N.~Ishibashi, M.~Li and A.R.~Steif\jo\prl&67(91)3336.}
\ref{C.~Holzhey, F.~Wilczek,``{\it Black Holes as Elementary Particles}'',
     IASSNS-HEP-91/71.}
\ref{G.~Gilbert,``{\it The Instability of String-Theoretic Black Holes}'',
     UMDEPP 92-110.}
\ref{E.Raiten,``{\it Perturbations of a Stringy Black Hole}'',
     FERMI-PUB-91/338-T.}
\ref{A.~Carlini, F.~Fucito and M.~Martellini,``{\it On the Stability of
     a Stringy Black Hole}'', Preprint,1992.}
\ref{M.D.~McGuidan. C.R.~Nappi and S.A.~Yost,``{\it Charged Black
    Holes in Two-Dimensional String Theory}'', IASSNS-HEP-91/57.}
\ref{H.C.~Ohanian,``{\it Gravitation and Spacetime}'',
     ( W.W.Norton and Company, Inc. New York 1977).}
\ref{S.~Chandrasekhar,``{\it The Mathematical Theory of Black Holes}'',
     (Clarendon Press, Oxford. 1983).}
\ref{K. Gottfried, ``{\it Quantum Mechanics  Vol 1: Fundamental}''
     , (Addison-Wesley Publishing Co. Inc. 1989).}
\ref{S.P.~de Alwis,``{\it Comments on No-Hair Theorems and
     Stability of Black Holes}'', Boulder preprint COLO-HEP-267.}
\refout
\endpage
\centerline {\bf FIGURE CAPTION }
Figure 1:
     The effective potential $V(r^*)$ {\it vs.}   coordinate $r^*$.
     The parameters chosen are $Q=1$, mass $m=1$ and for various
	  values of charge $q$.
\end
\bye